\documentclass[fleqn,usenatbib]{mnras}

\usepackage[T1]{fontenc}
\usepackage[utf8]{inputenc}
\usepackage{newunicodechar}
\usepackage{graphicx}
\usepackage{amsmath}
\usepackage{booktabs}

\newunicodechar{⊕}{\ensuremath{_\oplus}}
\newunicodechar{☉}{\ensuremath{_\odot}}
\newunicodechar{★}{\ensuremath{_\star}}
\newunicodechar{σ}{\ensuremath{\sigma}}
\newunicodechar{δ}{\ensuremath{\delta}}
\newunicodechar{Δ}{\ensuremath{\Delta}}
\newunicodechar{β}{\ensuremath{\beta}}
\newunicodechar{ω}{\ensuremath{\omega}}
\newunicodechar{π}{\ensuremath{\pi}}
\newunicodechar{ρ}{\ensuremath{\rho}}
\newunicodechar{χ}{\ensuremath{\chi}}
\newunicodechar{≈}{\ensuremath{\approx}}
\newunicodechar{≤}{\ensuremath{\leq}}
\newunicodechar{≥}{\ensuremath{\geq}}
\newunicodechar{±}{\ensuremath{\pm}}
\newunicodechar{×}{\ensuremath{\times}}
\newunicodechar{−}{\ensuremath{-}}
\newunicodechar{″}{\ensuremath{''}}
\newunicodechar{°}{\ensuremath{^\circ}}
\newunicodechar{†}{\ensuremath{^\dagger}}

\newcommand{\Pcat}{\ensuremath{P_{\rm cat}}}
\newcommand{\Rp}{\ensuremath{R_{\rm p}}}
\newcommand{\Teff}{\ensuremath{T_{\rm eff}}}

\title[Are habitable-zone TOIs really there?]{Are habitable-zone TESS Objects of Interest really there? Period resolution from the full archive}

\author[J. F. Pagés Navarrete]{Juan F. Pagés Navarrete\thanks{E-mail: jfpages@gmail.com}
\\
Independent researcher, Málaga, Spain. ORCID: 0009-0006-7910-4276}

\date{Accepted XXX. Received YYY; in original form ZZZ}
\pubyear{2026}

\begin{document}
\label{firstpage}
\pagerange{\pageref{firstpage}--\pageref{lastpage}}
\maketitle

\begin{abstract}
TESS Objects of Interest (TOIs) seen as single or duo transits carry a catalogue period that is undefined or equal to the separation of the two events, and their insolation and any habitable-zone label inherit it. We reanalyse all 187 TOIs catalogued with $P = 0$ or $P \geq 200$\,d using all public TESS data through Sector 107, with a flux-based method that weights each alias by per-epoch transit likelihoods, the transit duration and a geometric prior, and tests for coherence, data edges, contamination by other planets and instrumental artefacts. It recovers 12 published periods blindly and, in injection--recovery tests, 71 per cent of periods constrained by a new transit, with no wrong period. It resolves 28 periods and fixes 8 more by exclusion. Are habitable-zone TOIs really there? Mostly not: of the 28 duo-transits that a catalogue-period criterion places in the conservative habitable zone, about two are expected to lie there, both giant planets, and none of the 12 small ($\Rp < 4$\,R⊕) ones; those resolved are aliases of hotter planets. Conversely, habitable-zone planets hide behind missing or wrong periods: we recover three published M-dwarf examples and validate TOI-2298\,b, a 3.0\,R⊕ sub-Neptune at 146.58\,d in the conservative habitable zone of a K dwarf. Of three further planets validated with new periods, TOI-6075\,b (1.7\,R⊕, 416.5\,d, 833\,d not excluded) lies beyond the outer edge and is the second validated transiting planet smaller than 2\,R⊕ with $P > 1$\,yr. We release alias-marginalised habitable-zone probabilities for the whole sample.
\end{abstract}


\section{Introduction}

Transiting planets on orbits of months to years are scarce but disproportionately valuable. They probe the cool end of the sub-Neptune population, the region where atmospheres are expected to retain volatiles, and the habitable zones of Sun-like and K-type stars. Because the transit probability falls as $P^{-2/3}$ and the number of transits per observing window falls as $P^{-1}$, these planets are found at a low rate and are the hardest to characterise (see the recent review by \citealt{UlmerMoll2026}).

The Transiting Exoplanet Survey Satellite \citep[TESS;][]{Ricker2015} observes most of the sky in 27-d sectors and revisits fields at intervals of roughly two years. A planet with a period of a few months is therefore often seen once (a mono-transit) or twice in non-consecutive sectors (a duo-transit), and such objects were predicted to be numerous \citep{Villanueva2019,Cooke2019,Cooke2021}. For a duo-transit, the true period is one of a discrete set of aliases $\Delta/n$, where $\Delta$ is the separation between the two events; for a mono-transit, it is constrained only by the transit duration and by the absence of further events in the data. The TOI catalogue \citep{Guerrero2021} records such objects with $P = 0$ or with $P = \Delta$, the longest alias.

Resolving these aliases has motivated a growing body of work. Probabilistic alias ranking from TESS data and stellar density, as implemented in \textsc{MonoTools}, prioritises follow-up of the most likely periods \citep{Osborn2022}. Dedicated programmes have characterised samples of TESS duo-transits and pursued them from the ground and with CHEOPS \citep{Hawthorn2024}; the young sub-Neptunes of TOI-2076, whose outer periods had 11 and 7 allowed aliases, are a canonical example \citep{Osborn2022}. Individual periods have also been recovered from later TESS sectors, sometimes revealing that the catalogue value was a harmonic of the true one \citep{Fukui2025}. Most recently, uniform reanalyses of TOI host samples have resolved several long-period candidates around M dwarfs \citep{Tschudi2026a,Tschudi2026b}, and ground-based networks have confirmed long-period TESS planets and identified eclipsing binaries among them \citep{Rea2026}.

The closest precedent to this work is \citet{Bass2025}, who searched every available TESS reduction at the alias epochs of the 266 TOIs with catalogue periods longer than 27\,d released up to Sector 62 (code: \citealt{Bass2024}). They found unrecorded transits at aliases of eight TOIs, confirmed long periods for nine more, and concluded that most TOIs with catalogue periods above 500\,d are duo-transits with almost no data at their aliases. Their search excludes mono-transits, which have no catalogue period; it identifies transits by $\chi^2$ fits above a signal threshold rather than weighting the aliases probabilistically; and it predates most of the second extended mission. Since then the TESS archive has kept growing. By the end of the second extended mission many fields have been observed in four or more years, and a substantial number of catalogued duo- and mono-transits already have additional, unrecorded transits in public data. Their catalogue periods, and therefore their insolations, may be wrong. This matters beyond bookkeeping: lists of habitable-zone TOIs and demographic studies of long-period planets inherit these errors. The Revised TESS Habitable Zone Catalog \citep{Kaltenegger2021}, an update of \citet{Kaltenegger2019}, for example, recomputed the insolation of every TOI with TIC-8 stellar parameters but adopted the catalogue periods, and listed three TOIs in the habitable zone: TOI-203.01, TOI-715.01 and TOI-2298.01. The authors themselves noted that TOI-2298.01 had a single transit and a catalogued period of 165.2\,d, and that the 164-d catalogue period of TOI-793.01 was in fact 54\,d, too hot for the habitable zone. To our knowledge no habitable-zone list of TOIs has since been built that accounts for period aliases. Habitable-zone catalogues exist for confirmed planets \citep{Hill2023}, and the Kepler Habitable Zone Working Group catalogued the Kepler candidates in the habitable zone \citep{Kane2016}, whose periods were well determined by four years of continuous coverage. TESS has no equivalent maintained list of candidates, and its long-period TOIs, observed in separate sectors, are exactly those whose catalogue periods are least reliable.

In this paper we perform a systematic, full-archive reanalysis of every TOI catalogued with $P = 0$ or $P \geq 200$\,d, with two aims: to test whether TOIs catalogued in the habitable zone are really there, and to find habitable-zone planets hidden behind incorrect catalogue periods. We develop a flux-based alias analysis designed to be robust against the failure modes we encountered (systematics at data edges, contamination by other planets in the same system, and duos whose two events do not share a period), validate it blindly on published periods, and apply it to the full sample (Sections~\ref{sec:sample}--\ref{sec:census}). We then characterise and statistically validate four planets whose periods we determine for the first time, including a sub-Neptune in the conservative habitable zone of a K dwarf (Section~\ref{sec:planets}), and discuss the implications for TESS habitable-zone catalogues and for follow-up (Section~\ref{sec:discussion}). Relative to \citet{Bass2025}, we add mono-transits, use data through Sector 107, weight the aliases with the transit duration, the stellar density and a geometric prior in the spirit of \textsc{MonoTools} \citep{Osborn2022}, introduce coherence and contamination tests, validate the method blindly on published periods, and carry the strongest cases through statistical validation.

\section{Sample and data}
\label{sec:sample}

\subsection{Sample selection}

We downloaded the TOI catalogue from ExoFOP on 2026 October 21 and retained objects that (i) are not dispositioned as false positives (FP) or false alarms (FA), (ii) orbit hosts with $2600 \leq \Teff \leq 7200$\,K, the range of the \citet{Kopparapu2014} habitable-zone limits, and (iii) have either no catalogued period (mono-transits, $P = 0$) or $P \geq 200$\,d. The second criterion is a proxy for duo-transits, whose catalogue period is the separation $\Delta$ between two events. It misses duo-transits with $\Delta < 200$\,d and may include a few objects with three or more catalogued transits; we classify each object again from the data in Section~\ref{sec:trials}. The final sample contains 187 TOIs: 94 catalogued duo-transits and 93 mono-transits.

The selection was designed to isolate the TOIs whose catalogue period, and therefore whose insolation and habitable-zone status, is least reliable. TESS usually revisits a field after one or two years, so the separations of duo-transits cluster near 350 and 700\,d; most objects in our sample have catalogue periods of 700--750\,d. The 200-d threshold captures these revisit separations while excluding most objects with three or more transits, whose periods are already fixed by the data.

The same threshold matches the habitable-zone periods of the stars where the catalogue is most likely to mislead. For a Sun-like star the conservative habitable zone \citep{Kopparapu2014} corresponds to periods of about 340--790\,d, and for F stars to longer periods still; for a K dwarf such as TOI-2298 ($L = 0.15$\,L☉) it spans about 105--265\,d. A duo-transit with $\Pcat \geq 200$\,d around an FGK star is therefore often catalogued in or near the habitable zone, whatever its true period. Habitable-zone periods around M dwarfs (roughly 10--100\,d) are not selected directly, but they enter the sample in two ways: through mono-transits, which have no catalogue period, and as short aliases of long catalogue periods. The three M-dwarf planets we find in the habitable zone (TOI-6478\,b, TOI-6981\,b and TOI-2433.01) entered in exactly these ways. The criterion thus serves both aims of this work: to test whether catalogued habitable-zone candidates are real, and to find habitable-zone planets hidden behind incorrect catalogue periods.

\subsection{Light curves}

For each target we used all public TESS light curves available through MAST, accessed with \textsc{lightkurve} \citep{Lightkurve2018}: SPOC 120-s light curves \citep{Jenkins2016} when available, otherwise TESS-SPOC full-frame-image light curves \citep{Caldwell2020}, and otherwise QLP \citep{Huang2020,Kunimoto2021}, taking one product per sector in that order of priority. We used the PDCSAP (or equivalent) flux, which is corrected for crowding by the pipeline, applied the default quality mask, and normalised each sector. No global detrending was applied; all transit measurements use a local baseline (Section~\ref{sec:ses}). For each star we stored the full time series, which allows the analysis of every predicted epoch without new downloads.

\subsection{Stellar parameters}

Stellar effective temperatures, radii, surface gravities and masses for the alias analysis were taken from the ExoFOP catalogue, which reflects the TESS Input Catalog v8.2 \citep{Stassun2019}. For the final planets we adopted Gaia DR3 parameters \citep{Gaia2023}: luminosity and mass from FLAME and \Teff\ from GSP-Phot \citep{Creevey2023}, with percentile uncertainties, falling back to TIC values where FLAME is not available. We also retrieved the renormalised unit weight error \citep[RUWE;][]{Lindegren2021} and \texttt{ipd\_frac\_multi\_peak} as indicators of unresolved binarity.

\subsection{Ancillary data}
\label{sec:ancillary}

From ExoFOP we collected high-resolution imaging for the validated planets: WIYN/NESSI speckle contrast curves for TOI-2298 (832\,nm) and TOI-5724 \citep[562\,nm;][]{Scott2018}, a Gemini-South/Zorro curve at 832\,nm for TOI-4629 \citep{Scott2021}, and a SAI 2.5-m speckle curve in the $I$ band for TOI-6075 \citep{Strakhov2023}, which is available only as a figure and was digitised (27 points; precision ≈\,0.01″ in separation and ≈\,0.05\,mag in contrast). Reconnaissance spectra from FLWO/TRES exist for TOI-2298 (three epochs) and TOI-6075 (five epochs).

\section{Period resolution method}
\label{sec:method}

\begin{figure*}
\centering
\includegraphics[width=\textwidth]{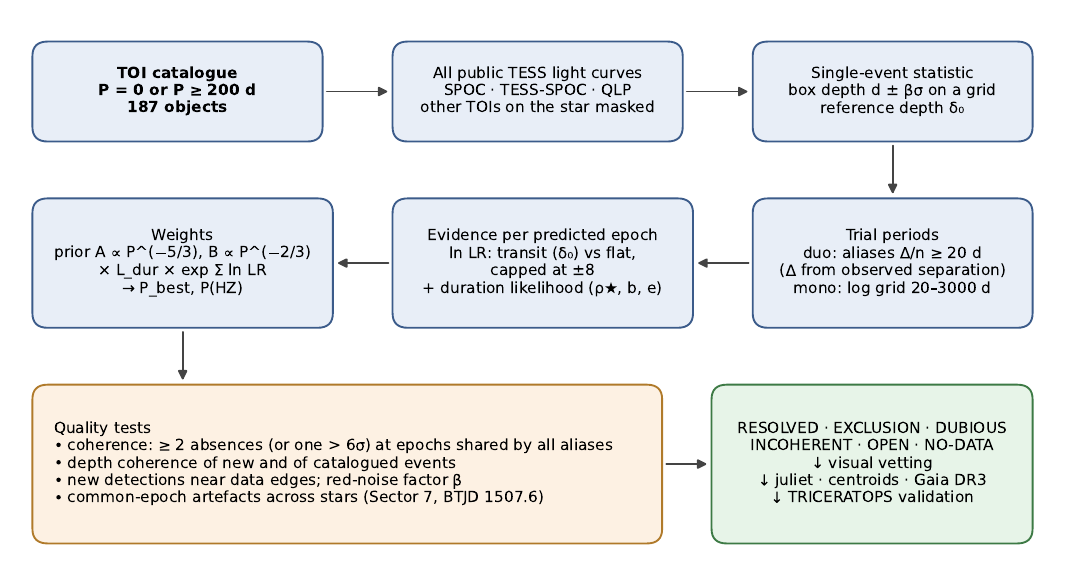}
\caption{Flow of the period-resolution method. For each TOI, the single-event statistic is evaluated at every epoch predicted by each trial period; the per-epoch evidence, the transit-duration likelihood and a geometric prior give the weight of each period; quality tests and visual vetting assign the final state, and the strongest cases proceed to transit fitting, centroid analysis, Gaia DR3 checks and statistical validation.}
\label{fig:method}
\end{figure*}

The method asks, for every trial period, whether the flux at each epoch it predicts looks like the catalogued transit, like no transit, or is uninformative, and combines the answers with the transit duration and a prior. Fig.~\ref{fig:method} summarises the steps.

\subsection{Single-event statistic}
\label{sec:ses}

For each star we compute a single-event statistic (SES; in the spirit of the Kepler and TESS pipelines, \citealt{Jenkins2002}) on a regular grid of trial mid-times with step $D/4$, where $D$ is the catalogued transit duration. At each grid time $t_c$ we select data within a window of $\pm3D$ ($\pm2D$ for transits longer than 12\,h) and fit a baseline to the out-of-transit points ($|t - t_c| > 0.75D$, at least three on each side), linear for windows shorter than 1\,d and quadratic otherwise to follow stellar variability, and measure the box depth $d$ as the mean baseline-subtracted deficit of points with $|t - t_c| < 0.35D$. The uncertainty $\sigma$ is 1.4826 times the median absolute deviation of the out-of-transit residuals divided by the square root of the number of in-transit points. To absorb correlated noise, all $\sigma$ are multiplied by a factor $\beta \geq 1$, the robust standard deviation of $d/\sigma$ over the whole grid. The reference depth $\delta_0$ is the inverse-variance weighted mean depth of the catalogued events detected at S/N\,$> 3$; unlike the median, it is not dragged by a single noisy outlier among two events.

\subsection{Trial periods}
\label{sec:trials}

For duo-transits we evaluate the SES at all catalogued epochs $T_0 + m\Pcat$ that fall on data and identify those with S/N\,$> 3$. The base separation is $\Delta = g\,\Pcat$, where $g$ is the greatest common divisor of the differences between detected epoch indices; $g > 1$ flags catalogue periods that are already harmonics of the observed separation (e.g. TOI-4600.02). The trial periods are the aliases $P_n = \Delta/n$ with $P_n \geq 20$\,d; shorter periods would have produced a second transit within a single sector. If only one catalogued event is visible, the object is treated as a mono-transit. For mono-transits the trial periods form a logarithmic grid between 20 and 3000\,d with step $\Delta\ln P = D/(3B)$, where $B$ is the time baseline of the data, so that predicted epochs drift by less than $D/3$ across the baseline.

\subsection{Per-epoch evidence}

For each trial period we predict all transit epochs that fall on the SES grid and compare, at each one, the hypothesis of a transit of depth $\delta_0$ with that of no transit:
\begin{equation}
\ln \mathrm{LR}_k = \ln \mathcal{N}\!\left(d_k \mid \delta_0, \sqrt{\sigma_k^2 + (s\,\delta_0)^2}\right) - \ln \mathcal{N}\!\left(d_k \mid 0, \sigma_k\right).
\end{equation}
The term $s\,\delta_0$, with $s = 0.3$, allows for depth differences between sectors and products (different dilution corrections in SPOC and QLP). Each $\ln\mathrm{LR}_k$ is capped at $\pm8$, so that no single epoch decides the outcome. Epochs predicted by every trial period (the catalogued events and, for duos, all multiples of $\Delta$) contribute equally to all trials and cancel on normalisation; they are handled separately by the coherence test (Section~\ref{sec:tests}).

\subsection{Transit duration and priors}

The duration likelihood $L_{\rm dur}(P)$ compares the observed duration with the model duration
\begin{equation}
T = 13\,\mathrm{h}\left(\frac{P}{365\,\mathrm{d}}\right)^{1/3}\left(\frac{\rho_\star}{\rho_\odot}\right)^{-1/3}\sqrt{1-b^2}\;\frac{\sqrt{1-e^2}}{1+e\sin\omega},
\end{equation}
marginalised by Monte Carlo over $b \sim \mathcal{U}(0, 1)$, $e \sim$ Rayleigh(0.15) truncated at 0.9, $\omega \sim \mathcal{U}(0, 2\pi)$ and a 20 per cent log-normal uncertainty on $\rho_\star$, with a Gaussian of width $\max(\sigma_D, 0.1D)$ around the observed duration. We use two priors to bracket prior sensitivity: prior A combines the geometric transit probability with an occurrence rate uniform in $\log P$; prior B uses the geometric probability alone. For the discrete aliases of a duo-transit these give weights proportional to $P^{-5/3}$ and $P^{-2/3}$; for the logarithmic grid of mono-transits, $P^{-2/3}$ and $P^{1/3}$ per cell. The posterior weight of each trial period is the product of the prior, $L_{\rm dur}$ and $\exp(\sum_k \ln \mathrm{LR}_k)$, normalised over all trials. Unless stated otherwise we quote prior A.

\subsection{Masking of other TOIs}
\label{sec:masking}

Before computing the SES we remove all cadences within $0.75D_j + 0.02$\,d of the predicted transits of every other TOI on the same star, using its catalogued ephemeris and regardless of disposition. Without this step, transits of inner planets can masquerade as detections of the outer candidate: before masking, TOI-1873.02 was `resolved' at 47.4\,d by events that were transits of TOI-1873\,b ($P = 7.0$\,d), with three clear absences at other predicted epochs.

\subsection{Quality tests and classification}
\label{sec:tests}

We classify each object with the following tests, applied to the best trial period $P_{\rm best}$ and its harmonic family (ratios 1/3, 1/2, 2/3, 1, 3/2, 2, 3):
\begin{enumerate}
\item \textit{Global coherence (duos).} An epoch common to all aliases ($T_0 + j\Delta$), other than the detected catalogued events themselves, shows a clear absence if $d < 0.3\delta_0$ and $(\delta_0 - d)/\sigma > 4$. If two or more such absences occur, or a single one exceeds $6\sigma$, no alias can explain the data and the object is INCOHERENT; a single weaker absence is recorded but does not classify the object. The test is not applied when the catalogued events themselves have incompatible depths (item iv), in which case the object is DUBIOUS. These thresholds were calibrated with the injection--recovery tests of Section~\ref{sec:validation}.
\item \textit{Resolution.} If $P_{\rm best}$ or its family carries more than 90 per cent of the weight and at least one new epoch (not common to all trials) has $\ln\mathrm{LR} > 2$, the period is RESOLVED.
\item \textit{Exclusion.} A duo with more than 90 per cent of the weight on one alias but no new detections is resolved by EXCLUSION: all shorter aliases are ruled out by absences.
\item \textit{Dubious.} A RESOLVED or EXCLUSION result is downgraded to DUBIOUS if any of the following holds: clear absences at predicted new epochs; new events with mean depth outside 0.6--1.6\,$\delta_0$ at more than $3\sigma$; $\beta > 4$; at least half of the new detections within 0.5\,d of a data edge; or catalogued events whose depths differ by more than a factor of two at more than $3\sigma$.
\item \textit{Open and no data.} Objects with no dominant period are OPEN; those whose catalogued event is not detected in the downloaded data are NO-DATA.
\end{enumerate}

\subsection{Common-epoch artefacts}
\label{sec:artefacts}

Instrumental events can masquerade as transits in several stars at once. Events of this kind are known to produce spurious long-period TOIs: \citet{Kaltenegger2021} rejected TOI-203.01, nominally in the habitable zone, because two of its four events coincided with pointing jumps from planned thruster firings. We therefore searched for catalogued events of different stars that fall within 0.3\,d of each other, considering both the catalogued epoch and, for duo-transits, the second event, and inspected every group of three or more stars. If catalogued epochs were independent and uniform in time (about 280 events over roughly 3000\,d), about 0.1 chance groups of three stars and $10^{-3}$ of four would be expected. They are not independent, however: TOIs alerted from the same pipeline search of the same data share its systematics, so a coincidence of epochs is a reason for suspicion rather than a proof.

We therefore test each suspected epoch directly against the other stars observed on the same detector, which have no reason to show a transit at that time. For every star with 2-min data on the same camera and CCD we fit a linear baseline to the quality-filtered flux within 1\,d of the epoch, excluding the transit window, and measure the mean deficit in the window and its significance $z$ relative to the white-noise scatter of the residuals. We do this in both the simple-aperture (SAP) flux and the PDCSAP flux, because PDC cotrending is designed to remove signals common to many stars, and compare the fraction of stars with $z > 3$ at the suspected epoch with its distribution over a grid of epochs spanning the sector. The comparison with other epochs absorbs the effect of correlated noise, which makes $z > 3$ far more common than for white noise.

\begin{figure}
\centering
\includegraphics[width=\columnwidth]{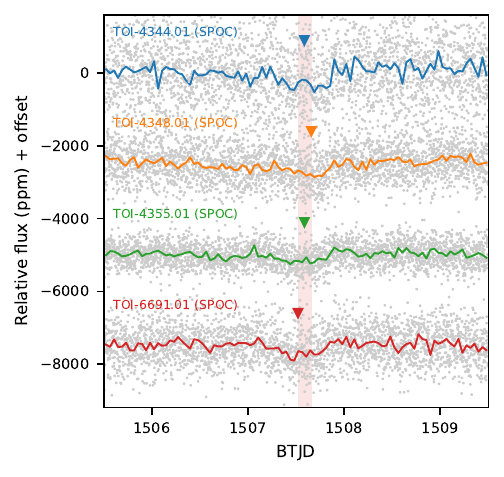}
\caption{SPOC 2-min light curves of TOI-4344.01, TOI-4348.01, TOI-4355.01 and TOI-6691.01 around BTJD 1507.6 (Sector 7), binned to 1\,h and vertically offset. Triangles mark each catalogued epoch and the shaded band the interval BTJD 1507.52--1507.66 that contains all four. The simultaneous shallow dip in four unrelated stars, shared by the other stars on the same detector (Fig.~\ref{fig:commonmode}), identifies an instrumental artefact.}
\label{fig:sector7}
\end{figure}

\begin{figure}
\centering
\includegraphics[width=\columnwidth]{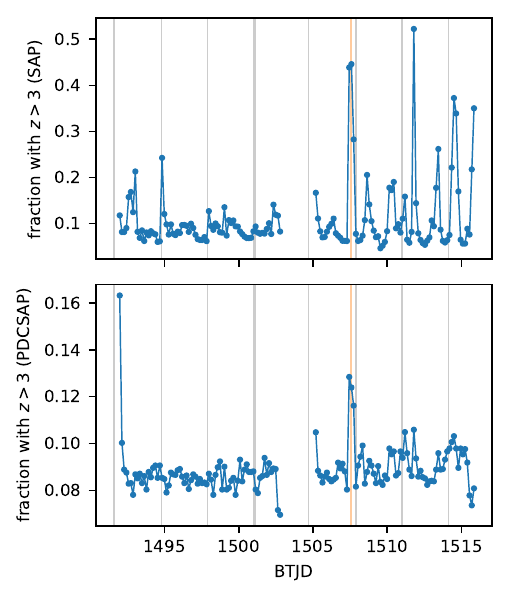}
\caption{Common-mode test for the Sector 7 artefact. Fraction of the 3995 non-TOI stars with 2-min data on Camera 4, CCDs 2 and 3, that show a dip with $z > 3$ (white-noise significance) in a 3.4-h window, as a function of the window centre, in SAP (top) and PDCSAP (bottom) flux. The shaded band marks the window BTJD 1507.52--1507.66 that contains the four catalogued events, and grey lines the momentum dumps. The baseline level of about 8 per cent reflects correlated noise.}
\label{fig:commonmode}
\end{figure}

One group stands out: TOI-4344.01, TOI-4348.01, TOI-4355.01 and TOI-6691.01 have catalogued events between BTJD 1507.52 and 1507.66, within three hours, in Sector 7. Three of them were alerted on the same day from SPOC multi-sector searches of the same data, and all four light curves show a shallow simultaneous dip at that time (Fig.~\ref{fig:sector7}). All four stars lie on Camera 4 (three on CCD 2, one on CCD 3). In SAP flux each shows a dip of 440--580\,ppm (7--13$\sigma$) in the window, about twice as deep as in PDCSAP (180--265\,ppm), some 6\,h before a momentum dump at BTJD 1507.86. Among the 3995 other stars with 2-min data on those two CCDs, 45 per cent show a SAP dip with $z > 3$ in the same window (median depth 360\,ppm), against a median of 8 per cent over the sector, with the same result on both CCDs (Fig.~\ref{fig:commonmode}). PDC removes the dip for most stars (median ≈\,0\,ppm) but leaves an excess of stars with residual dips (12.4 against 8.7 per cent). Apart from the first day of the sector, the 1507.5 window is the only epoch of Sector 7 with such an excess in PDCSAP, the flux searched by SPOC; similar SAP features elsewhere in the sector, the strongest near BTJD 1511.8, are removed by PDC. The four TOIs are therefore stars in which PDC only partly removed a camera-wide common-mode event, and we exclude them from the results. The same test identifies the first catalogued event of TOI-4555.01 as a systematic of the same kind (Section~\ref{sec:validation}). A second group of three (TOI-4361.01, TOI-4827.01 and TOI-6492.01, BTJD 2261.98--2262.12) is consistent with chance: the event of TOI-6492.01 is a deep transit of a published planet, and that of TOI-4361.01 is consistent with its resolved period.

\subsection{Visual vetting}

Every RESOLVED, EXCLUSION and DUBIOUS object was inspected by eye using per-event panels (individual transits marked as catalogued or new, with depth, S/N, optimal box width and flags for nearby gaps and data edges) and a phase-folded light curve, together with odd--even depths, the depth at phase 0.5 and, for exclusions, the measured depth at every epoch predicted by the aliases $\Delta/2$ to $\Delta/5$. The automatic classification and the visual verdict agree in most cases; we report both, and the final results rely on the visual verdict.

\section{Validation of the method on published periods}
\label{sec:validation}

\begin{figure}
\centering
\includegraphics[width=\columnwidth]{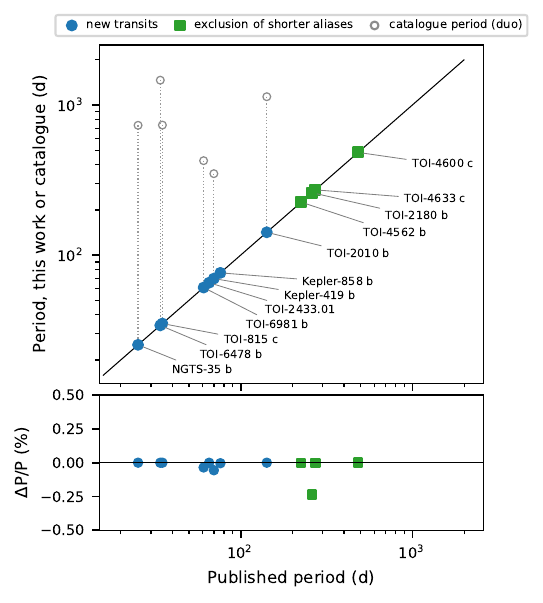}
\caption{Blind recovery of 12 published periods. Top: period found in this work against the published value (filled symbols; circles for resolutions by new transits, squares for exclusion of all shorter aliases); open circles mark the catalogue period of duo-transits, joined to the true period by dotted lines. Bottom: fractional difference. The $-0.24$ per cent offset is TOI-2180\,b (260.17\,d here, 260.79\,d published).}
\label{fig:validation}
\end{figure}

The method recovers 12 published periods without using any information beyond the TOI catalogue and the TESS light curves (Table~\ref{tab:published}, Fig.~\ref{fig:validation}). We cross-matched all RESOLVED, EXCLUSION and DUBIOUS objects by TIC with the NASA Exoplanet Archive composite planet table and searched the literature for periods not yet in the archive.

Ten periods match confirmed planets in the archive to better than 0.5 per cent: TOI-6478\,b \citep[34.005\,d;][]{ScottM2025}, TOI-815\,c (34.976\,d), TOI-2010\,b (141.834\,d), NGTS-35\,b (25.241\,d), Kepler-419\,b \citep[69.716\,d against 69.755\,d;][]{Dawson2014}, Kepler-858\,b \citep[76.138\,d;][]{Morton2016}, TOI-4562\,b (225.104\,d), TOI-4633\,c (271.94\,d), TOI-4600\,c (482.82\,d) and TOI-2180\,b \citep[260.17\,d against 260.79\,d;][]{Dalba2022}. Two more match planets published after the archive snapshot: TOI-6981\,b at 60.78\,d \citep{ScottM2026} and TOI-2433.01 at 65.459\,d \citep{Tschudi2026b}. Eight of the twelve were resolved by new transits and four by exclusion of all shorter aliases (TOI-2180.01, TOI-4562.01, TOI-4600.02 and TOI-4633.01), in which case the catalogued separation is the true period.

Three further cases test how the method fails. TOI-4353.01, resolved at 39.899\,d by the same M-dwarf survey \citep{Tschudi2026b}, is our highest-weight period ($w = 0.81$) but falls below the 0.90 threshold because of one clear absence at a predicted epoch; the method ranks it correctly but does not claim it. TOI-6492.01 \citep[published at 33.265\,d;][]{Tschudi2026b} is left OPEN: all its TESS events in our data lie at multiples of the catalogue separation, so the data contain no information on the alias, and the method correctly declines to choose; \citet{Tschudi2026b} resolved it with two consecutive transits including a later sector. TOI-4361.01 is resolved at 26.479\,d ($n = 28$), inside the range of aliases allowed by an independent analysis \citep[between $\Delta/34$ and $\Delta/21$;][]{Mantovan2022}.

An independent uniform analysis of 461 M-dwarf TOI hosts \citep{Tschudi2026b} allows a further comparison. Both analyses recover TOI-2433.01 (65.459\,d) and TOI-4353.01 (39.899\,d, our highest-weight period), and both leave TOI-4349.01 unresolved. Tschudi resolves TOI-6492.01 with a transit in a sector later than our data, and TOI-4565.01, which he finds at 19.786\,d, is outside our sample because its catalogue period had already been revised to that value.

Three failure modes were identified during development and motivated the quality tests of Section~\ref{sec:tests}. First, a hard veto that rejects any period predicting a transit on data excluded the true period of TOI-4353.01, TOI-6492.01 and TOI-2433.01, because the catalogue did not record transits that are present in later sectors; the soft, flux-based veto fixes this. Second, the cancellation of epochs common to all aliases allowed TOI-4555.01 and TOI-2090.01 to be resolved by exclusion even though their own catalogued separation predicts a transit that is absent (for TOI-4555.01, $-17 \pm 48$\,ppm against an expected 305\,ppm, $6.7\sigma$); the global coherence test now classifies them as INCOHERENT. Third, transits of other planets in the same system produced spurious detections (TOI-1873.02, Section~\ref{sec:masking}), which masking removes while leaving genuine resolutions such as TOI-815\,c unchanged.

Our results also reproduce the new aliases that \citet{Bass2025} reported for targets common to both samples: TOI-4633.01 at 271.94\,d and TOI-5975.01 at 186.52\,d, and both analyses find no data at the 369.6-d alias of TOI-1894.01. The two analyses disagree on four objects, all of which \citet{Bass2025} flag as low-signal candidates (their table~3). TOI-4356.01, for which they report a new alias at 289.82\,d, remains OPEN in our final run because only one of its catalogued events is detected at S/N\,$> 3$ in the data we use; this is a non-detection rather than a contradiction, and the TOI catalogue comments also note a low S/N and a poor transit shape. For TOI-4555.01, our global-coherence test finds a clear absence at an epoch predicted by every alias (BTJD 3091.7, Sector 65; $-17 \pm 48$\,ppm against an expected 305\,ppm, $6.7\sigma$), so no single period explains both catalogued events. This epoch lies after the Sector 62 limit of their data, so the disagreement reflects new data rather than a different reading of the same light curves. The absence is clean: the epoch is more than a day from any data gap, has no quality flags or momentum dumps, and a simple local measurement gives $-23 \pm 23$\,ppm in both SAP and PDCSAP flux. The first catalogued event (BTJD 1523.85, Sector 8, Camera 4), in contrast, begins at a momentum dump and is twice as deep in SAP as in PDCSAP flux (577 against 312\,ppm). Applying the test of Section~\ref{sec:artefacts} to 500 stars on the same CCD, 83 per cent show a SAP dip with $z > 3$ in that window (median depth 630\,ppm), the highest fraction of any epoch in the sector, against a median of 17 per cent (higher than in Sector 7 because the 9.9-h window is longer); in PDCSAP the excess shrinks to 32 against 14 per cent. TOI-4555.01 therefore most likely rests on one instrumental event and one weak event (Sector 27, ≈\,190\,ppm), rather than on two transits of one planet. TOI-4348.01 and TOI-4355.01 have their first catalogued event in the Sector 7 artefact (Section~\ref{sec:artefacts}), and we do not consider either a credible long-period candidate.

\begin{figure*}
\centering
\includegraphics[width=\textwidth]{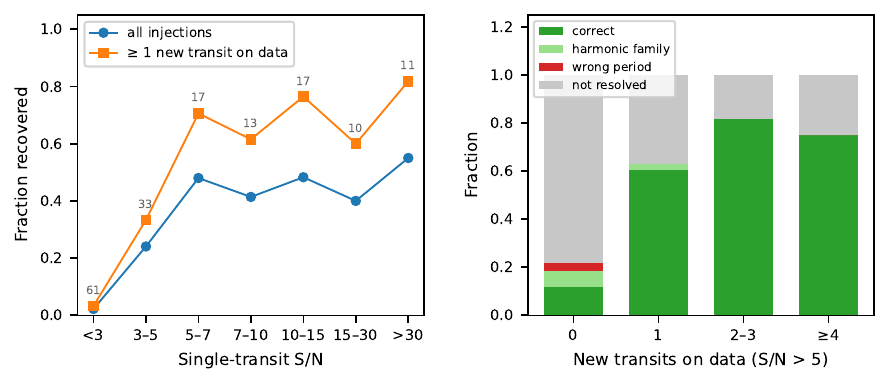}
\caption{Injection--recovery tests on 300 synthetic duo-transits. Left: fraction of injected periods recovered (correct period or harmonic family) as a function of the single-transit S/N, for all injections and for those with at least one transit beyond the two catalogued events on the data; numbers give the injections per bin. Right: outcome as a function of the number of new transits on the data, for S/N\,$> 5$.}
\label{fig:injection}
\end{figure*}

\textit{Injection--recovery tests.} To measure completeness and reliability we injected 300 synthetic transits into the real light curves of randomly chosen sample stars, after masking their own catalogued events. Periods were drawn log-uniformly between 20 and 800\,d, depths log-uniformly between 150 and 5000\,ppm, impact parameters uniformly in [0, 0.9], and durations from the stellar density. For each injection we built the object the catalogue would have recorded: a duo-transit with \Pcat\ equal to the separation of the first two observed transits when these are more than 200\,d apart, and discarded injections whose first two transits are closer, since the catalogue would have recorded their period. Each object was then analysed with the same code as the real sample. The results are summarised in Fig.~\ref{fig:injection}.

When at least one transit beyond the two catalogued events falls on the data and the single-transit S/N exceeds 5, the method recovers the injected period or a member of its harmonic family in 71 per cent of cases (74 per cent above S/N\,=\,10), leaves 12 per cent DUBIOUS and 18 per cent OPEN, and never returns a wrong period (0 of 68). These tests use duo-transits only; injections of mono-transits, whose continuous period grid is far more expensive to analyse, are left for future work. Over all 300 injections, 76 were classified as RESOLVED or EXCLUSION, of which 2 (2.6 per cent) had a wrong period. Both had no injected transit beyond the two catalogued events, so the method resolved on spurious events; this is the failure mode that visual vetting is designed to catch. One injection (0.3 per cent), with a single-transit S/N of 0.2, was classified as INCOHERENT. An earlier version of the coherence test, which compared a median $\delta_0$ with a single absence measured with a linear baseline over $\pm3D$, produced five false incoherences, four of them at S/N\,$> 7$; the corrections described in Sections~\ref{sec:ses} and~\ref{sec:tests} remove all of these.

\begin{table*}
\centering
\caption{Published periods recovered blindly. Not listed: the candidate TOI-5975.01, recovered at 186.519\,d, an alias first reported by \citet{Bass2025}.}
\label{tab:published}
\begin{tabular}{lllllll}
\toprule
TOI & Planet & \Pcat\ (d) & $P$, this work (d) & $P$, published (d) & Resolved by & Reference \\
\midrule
6478.01 & TOI-6478\,b & 1462.212 & 34.0049 & 34.005019 & new transit & \citet{ScottM2025} \\
6981.01 & TOI-6981\,b & 425.451 & 60.779 & 60.8 & new transit & \citet{ScottM2026} \\
2433.01 & -- & mono & 65.459 & 65.459 & new transits & \citet{Tschudi2026b} \\
815.02 & TOI-815\,c & 734.499 & 34.976 & 34.9761 & new transit & \citet{Psaridi2024} \\
2010.01 & TOI-2010\,b & 1134.672 & 141.834 & 141.834 & new transit & \citet{Mann2023} \\
6669.01 & NGTS-35\,b & 731.997 & 25.241 & 25.2412 & new transit & \citet{Kendall2026} \\
7909.01 & Kepler-419\,b & 348.579 & 69.716 & 69.755 & new transit & \citet{Dawson2014} \\
6083.01 & Kepler-858\,b & mono & 76.138 & 76.136 & new transits & \citet{Morton2016} \\
2180.01 & TOI-2180\,b & 260.174 & 260.174 & 260.79 & exclusion & \citet{Dalba2022} \\
4562.01 & TOI-4562\,b & 225.104 & 225.104 & 225.1064 & exclusion & \citet{Heitzmann2023} \\
4600.02 & TOI-4600\,c & 482.819 & 482.819 & 482.8191 & exclusion & \citet{Mireles2023} \\
4633.01 & TOI-4633\,c & 271.939 & 271.939 & 271.9445 & exclusion & \citet{Eisner2024} \\
\bottomrule
\end{tabular}
\end{table*}

\section{Census results and the habitable-zone correction}
\label{sec:census}

\subsection{Classification of the sample}

\begin{figure*}
\centering
\includegraphics[width=\textwidth]{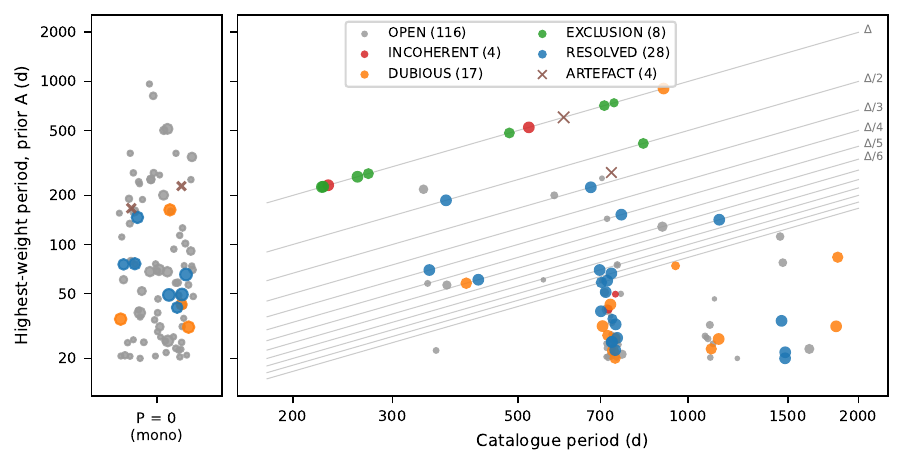}
\caption{Catalogue period against the highest-weight period (prior A) for the 187 TOIs, coloured by final state; symbol size scales with the weight of the best period. Grey lines mark the aliases $\Delta/n$ of the catalogue separation. Mono-transits, with no catalogue period, are shown in the left panel with random horizontal offsets. Crosses mark the four Sector 7 artefacts.}
\label{fig:census}
\end{figure*}

The automatic classification of the 187 TOIs gives 29 RESOLVED, 9 resolved by EXCLUSION, 17 DUBIOUS, 4 INCOHERENT (TOI-2090.01, TOI-4298.01, TOI-4555.01 and TOI-4563.01), 118 OPEN and 10 with no detectable catalogued event in the downloaded data (Table~\ref{tab:catalogue}, full version online; Fig.~\ref{fig:census}). Removing the two Sector 7 artefacts among them (TOI-4344.01, RESOLVED, and TOI-4348.01, EXCLUSION; Section~\ref{sec:artefacts}) leaves 28 resolved periods and 8 exclusions. Twelve catalogued duo-transits have only one event visible at S/N\,$> 3$ and were analysed as mono-transits. The fraction of resolved objects is much higher for duo-transits, whose discrete aliases are easily tested, than for mono-transits, whose periods are only weakly constrained unless the field has been observed repeatedly.

\begin{table*}
\centering
\caption{Alias-marginalised period and habitable-zone probability for the long-period TOIs discussed in the text; the full table of 187 objects is available online and as an arXiv ancillary file. Prior A; $\Pcat = 0$ for mono-transits. `Type' is the type as analysed: twelve catalogued duo-transits with a single detected event were analysed as mono-transits, and the full table also gives the type in the catalogue. Radii and insolations use TIC-8 parameters, as throughout the census; Section~\ref{sec:planets} and Table~\ref{tab:fits} give the values for the validated planets from the joint fits and Gaia DR3. The online version also gives a status summary for every object, prior-B values, harmonic-family weights, the catalogue insolation listed by ExoFOP (for mono-transits inherited from SPOC Data Validation reports; Section~\ref{sec:census}), the numbers of new and absent events at the best period, and notes.}
\label{tab:catalogue}
\small
\setlength{\tabcolsep}{3.5pt}
\begin{tabular}{lllrrrrrrrl}
\toprule
TOI & Type & State & \Pcat\ (d) & $P_{\rm best}$ (d) & $w_{\rm best}$ & $S$ (S⊕) & \Rp\ (R⊕) & $P_{\rm HZ,cons}$ & $P_{\rm HZ,opt}$ & Status \\
\midrule
2298.01 & mono & RESOLVED & 0 & 146.5788 & 0.936 & 0.565 & 2.90 & 0.982 & 0.982 & validated \\
5724.01 & duo & RESOLVED & 697.404 & 69.7404 & 1 & 19.267 & 3.02 & 0 & 0 & validated \\
4629.01 & duo & RESOLVED & 714.697 & 51.0498 & 1 & 23.854 & 2.49 & 0 & 0 & validated \\
6075.01 & duo & EXCLUSION & 832.924 & 416.4618 & 0.832 & 0.162 & 1.74 & 0.037 & 0.037 & validated \\
4503.01 & duo & EXCLUSION & 710.698 & 710.6982 & 0.815 & 0.201 & 2.09 & 0.185 & 0.185 & candidate \\
6478.01 & duo & RESOLVED & 1462.211 & 34.0049 & 0.996 & 0.383 & 4.54 & 0.998 & 0.998 & published \\
6981.01 & duo & RESOLVED & 425.451 & 60.7787 & 1 & 0.233 & 2.08 & 0 & 1 & published \\
2433.01 & mono & RESOLVED & 0 & 65.4588 & 1 & 0.239 & 1.81 & 0 & 1 & published \\
4600.02 & duo & EXCLUSION & 482.819 & 482.8191 & 0.836 & 0.316 & 6.91 & 0.837 & 0.837 & published, giant \\
1894.01 & duo & EXCLUSION & 739.239 & 739.2390 & 0.549 & 0.401 & 6.29 & 1 & 1 & giant; $P$ ambiguous \\
4349.01 & duo & OPEN & 749.160 & 74.9161 & 0.28 & 0.726 & 2.34 & 0.726 & 0.792 & HZ priority \\
4555.01 & duo & INCOHERENT & 522.627 & 522.6273 & 0.987 & 1.460 & 2.24 & -- & -- & 1st event instrumental \\
4356.01 & mono & OPEN & 579.647 & 200.1877 & 0.341 & 8.663 & 4.05 & 0.089 & 0.112 & low signal \\
4344.01 & mono & ARTEFACT & 731.481 & 276.2030 & 0.739 & 6.237 & 3.25 & -- & -- & artefact (S7) \\
4348.01 & duo & ARTEFACT & 602.245 & 602.2451 & 0.912 & 2.050 & 2.87 & -- & -- & artefact (S7) \\
4355.01 & mono & ARTEFACT & 0 & 228.1108 & 0.413 & 7.196 & 3.07 & -- & -- & artefact (S7) \\
6691.01 & mono & ARTEFACT & 0 & 166.8009 & 0.471 & 18.063 & 2.95 & -- & -- & artefact (S7) \\
\bottomrule
\end{tabular}
\end{table*}

Visual vetting of the 55 RESOLVED, EXCLUSION and DUBIOUS objects yields 12 published periods (Section~\ref{sec:validation}), one alias first reported by \citet{Bass2025} (TOI-5975.01), 7 new periods for small planets ($\Rp < 4$\,R⊕), 8 new periods for larger objects, one object with two allowed solutions (TOI-1894.01, 370 or 739\,d), 11 probable but visually weak resolutions and 15 rejections. The rejections fall into four groups: systematics at data edges or gaps posing as new transits (e.g. TOI-1563.01), new events with depths incompatible with the catalogued one (e.g. TOI-2472.01, TOI-4326.01), evident eclipsing binaries with V-shaped events tens of per cent deep (TOI-224.01, TOI-1379.01), and the two Sector 7 artefacts (TOI-4344.01 and TOI-4348.01; Section~\ref{sec:artefacts}). Among the larger objects, TOI-706.01 and TOI-4958.01 have secure periods but are probably not planets. All counts refer to the TOI catalogue downloaded from ExoFOP on 2026 October 21.

Two of the seven new small-planet periods are compromised by contamination. For TOI-4191.01 ($P = 22.511$\,d) the difference-image centroid is displaced by 11--16″ in the same direction in four sectors observed years apart, towards a Gaia source 1.4 pixels away. For TOI-7581.01 ($P = 226.28$\,d) two of five events are displaced, Gaia reports RUWE\,=\,2.4 and \texttt{ipd\_frac\_multi\_peak}\,=\,67, and the light curve contains deep, unrelated eclipses. We report their periods but do not claim the signals originate on the targets.

\subsection{The habitable-zone correction}

\begin{figure*}
\centering
\includegraphics[width=0.62\textwidth]{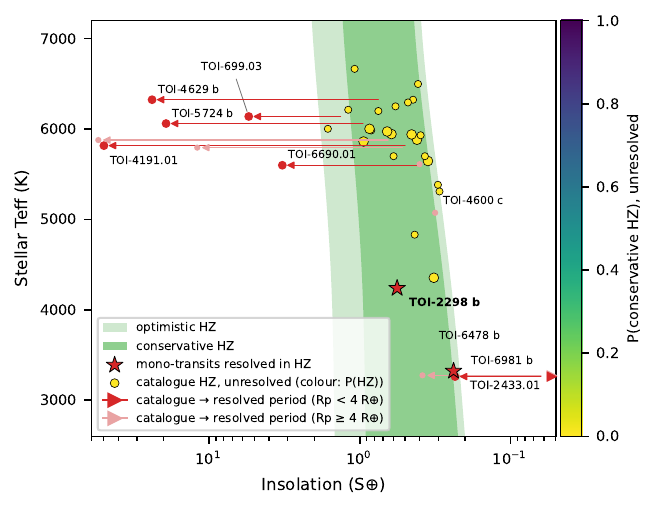}
\caption{Habitable-zone status of the long-period TOIs. Shaded bands show the conservative (dark) and optimistic (light) habitable zones of \citet{Kopparapu2013,Kopparapu2014}. Yellow circles are duo-transits that the catalogue period places in the habitable zone and that remain unresolved, coloured by their probability of lying in the conservative habitable zone after marginalising the aliases; all have P(HZ)\,≈\,0. Arrows go from the catalogue-period insolation to that of the resolved period (red for $\Rp < 4$\,R⊕, pink for larger objects); triangles at the right edge mark catalogue insolations below the plotted range. Stars are mono-transits resolved in the habitable zone. Insolations use TIC parameters (0.56\,S⊕ for TOI-2298\,b; 0.62\,S⊕ with Gaia DR3).}
\label{fig:hz}
\end{figure*}

The catalogue misplaces long-period TOIs relative to the habitable zone in both directions (Fig.~\ref{fig:hz}). Duo-transits catalogued in the habitable zone of FGK stars have periods of 700--900\,d, equal to the separation between visits two years apart. Whenever the full archive resolves them, the true period is a short alias and the planet is far hotter: TOI-4191.01 (53\,S⊕), TOI-4629.01 (20\,S⊕) and TOI-5724.01 (19\,S⊕) with Gaia DR3 parameters, and TOI-4361.01 and TOI-4326.01 (≈\,34--52\,S⊕ with TIC parameters). Conversely, duo- and mono-transits around M dwarfs that the catalogue places far beyond the habitable zone can have short aliases inside it: TOI-6478\,b (0.38\,S⊕), TOI-6981\,b (0.23\,S⊕) and TOI-2433.01 (0.24\,S⊕) were catalogued at 425--1462\,d or with no period. All three have since been published by other authors (Table~\ref{tab:published}); the only habitable-zone planet whose period is determined here for the first time is TOI-2298\,b (Section~\ref{sec:planets}).

For every object not classified as INCOHERENT or NO-DATA, and excluding the Sector 7 artefacts, we compute the probability of lying in the habitable zone, P(HZ), by summing the weights of all trial periods whose insolation falls within the habitable-zone limits at the host \Teff. Both sets of limits we use come from the climate models of \citet{Kopparapu2013,Kopparapu2014}: the conservative zone is bounded by the runaway greenhouse and maximum greenhouse limits (we adopt the 2014 coefficients for a one-Earth-mass planet, 1.11 and 0.36\,S⊕ for the Sun), and the optimistic or empirical zone by the recent Venus and early Mars limits (1.78 and 0.32\,S⊕ for the Sun). \citet{Kaltenegger2021} applied the latter, with the 2013 coefficients, to the TOIs. We then compare, for the same objects, the number that the catalogue period places in the habitable zone with the number expected once the period aliases are marginalised (Table~\ref{tab:hz}). We treat duo- and mono-transits separately, because the answer is robust for the former and prior-dependent for the latter.

\begin{table*}
\centering
\caption{Habitable-zone occupancy. For duo-transits, `Catalogue' counts the objects placed in the habitable zone by the criterion of \citet{Kaltenegger2021}, i.e. the insolation computed from the catalogue period and TIC-8 parameters; `Remain' counts those resolved here (RESOLVED or EXCLUSION) whose true period keeps them in the habitable zone; `Expected' is the sum of P(HZ) over the catalogue habitable-zone objects (priors B--A). Mono-transits cannot be classified by a catalogue-period criterion; for them only the expected number over the whole group is given, an upper envelope dominated by the prior. Sector 7 artefacts, INCOHERENT and NO-DATA objects excluded.}
\label{tab:hz}
\begin{tabular}{llccccc}
\toprule
 & & \multicolumn{4}{c}{Duo-transits} & Mono-transits \\
Limits & Size & Catalogue & Resolved & Remain & Expected & Expected, whole group \\
\midrule
Conservative & all & 28 & 9 & 2 (TOI-1894.01, TOI-4600.02) & 1.8--1.9 & 8.6--12.6 \\
Conservative & $\Rp < 4$\,R⊕ & 12 & 4 & 0 & 0.0--0.1 & 4.2--4.8 \\
Optimistic & all & 33 & 10 & 2 (TOI-1894.01, TOI-4600.02) & 1.9 & 17.0--22.8 \\
Optimistic & $\Rp < 4$\,R⊕ & 13 & 5 & 0 & 0.0--0.1 & 8.2--9.1 \\
\bottomrule
\end{tabular}
\end{table*}

\subsection{Duo-transits: the answer to the title question}

\begin{figure*}
\centering
\includegraphics[width=\textwidth]{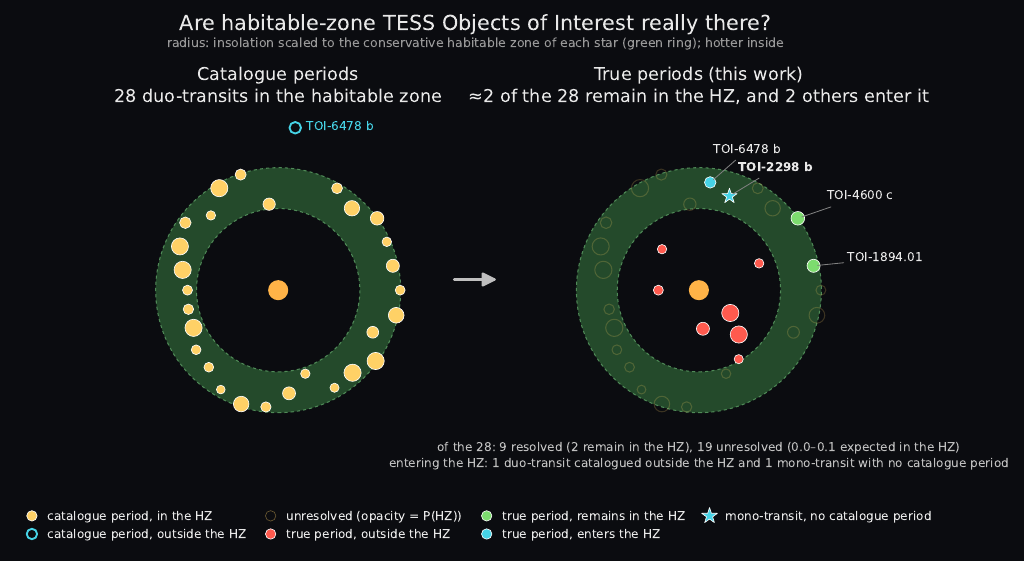}
\caption{The habitable-zone correction at a glance. Each symbol is a TOI placed at a distance from the centre that scales with its insolation relative to the conservative habitable zone of its host (green ring; hotter planets inside, colder outside); symbol size scales with planet radius. Left: positions from the catalogue periods, which place 28 duo-transits in the conservative habitable zone. Right: positions from the periods of this work. Of the 28, nine are resolved and only two remain in the habitable zone, both giant planets (TOI-4600\,c and TOI-1894.01); the 19 unresolved objects, drawn with opacity proportional to their probability of lying in the habitable zone, are expected to contribute 0.0--0.1. Two planets enter the habitable zone: TOI-6478\,b, a duo-transit catalogued outside it, and TOI-2298\,b, a mono-transit with no catalogue period.}
\label{fig:summary}
\end{figure*}

To define which duo-transits the catalogue places in the habitable zone we reproduce the criterion of \citet{Kaltenegger2021}: the insolation is computed from the catalogue period and the TIC-8 stellar parameters, and compared with the habitable-zone limits at the host \Teff. With the conservative limits this criterion places 28 duo-transits of our sample in the habitable zone, 12 of them small ($\Rp < 4$\,R⊕). We resolve the period of 9 of the 28; only two remain in the habitable zone, and both are giant planets: TOI-4600\,c (6.9\,R⊕, published) and TOI-1894.01 (6.3\,R⊕, likely in a binary and with its period still ambiguous between 370 and 739\,d). Summing P(HZ) over the 28 objects, only 1.8--1.9 are expected to lie in the habitable zone (priors B and A), and among the 12 small ones 0.0--0.1: none of the small duo-transits that a catalogue-period criterion places in the conservative habitable zone is expected to be there (Fig.~\ref{fig:summary}). The optimistic limits, equivalent to the empirical limits used by \citet{Kaltenegger2021}, give the same picture (33\,$\rightarrow$\,1.9; 13 small\,$\rightarrow$\,0.0--0.1). Using the catalogue insolation column instead of recomputing the insolation leads to the same conclusion (24 duo-transits in the conservative zone, of which 1.9--2.1 are expected to lie there). The result barely depends on the prior because duo-transit aliases are discrete and most of them are excluded or weighted down by the data. Every small catalogued habitable-zone duo-transit that the archive resolves (TOI-4191.01, TOI-4629.01 and TOI-5724.01, and tentatively TOI-699.03 and TOI-6690.01, whose resolutions are visually weak) turns out to be a short alias of a hotter planet.

To answer the question for the whole catalogue, we applied the same criterion to every TOI not dispositioned as a false positive or false alarm. Catalogue periods place 43 TOIs in the conservative habitable zone (59 in the optimistic one). Thirty-two of the 43 have $P \geq 200$\,d, 29 of them around FG stars, and all 32 are in our sample: the 28 analysed above, the INCOHERENT TOI-4298.01, and three without usable data (TOI-791.02, TOI-1895.01 and TOI-7060.02). The first of these illustrates the problem independently: catalogued at 928\,d, TOI-791.02 has been confirmed by \citet{Dransfield2026} as TOI-791\,c at 232.016\,d, a quarter of that value, receiving 4.1\,S⊕, well inside the inner edge of the habitable zone. The number of long-period TOIs genuinely in the conservative habitable zone therefore remains about two, both giant planets. The remaining 11 have $P < 200$\,d, and 10 of them orbit M dwarfs. Their periods are secure: those with fewer than three expected transits in the TESS data are all confirmed or known planets (e.g. TOI-256.01, LHS\,1140\,b), and the unconfirmed ones have three or more. The alias problem is therefore confined to the long-period TOIs, which make up three quarters of the catalogue's habitable-zone candidates and almost all of those around Sun-like stars. Two objects flagged by \citet{Kaltenegger2021} illustrate it: TOI-1823.01, then catalogued at 194\,d, is now a confirmed planet at 38.81\,d, one fifth of that period; and TOI-793.01, catalogued at 164\,d, is now a false positive at 54.66\,d.

The effect also works in the other direction. Summed over all duo-transits, the expected number in the conservative habitable zone is 6.3--6.8, of which only about two come from objects the catalogue lists there. The remainder comes from duo-transits that the catalogue places outside the habitable zone, mainly M-dwarf candidates whose shorter aliases fall inside it: the published planets TOI-6478\,b and TOI-6981\,b (the latter close to the outer edge), and unresolved candidates such as TOI-4349.01. With no new transits or absences in the data, TOI-4349.01 (2.3\,R⊕, $\Teff = 3851$\,K, catalogued at 749\,d) cannot be resolved, yet its probability of lying in the conservative habitable zone is 0.73 under both priors, because most of its allowed aliases fall inside the habitable zone of its early-M host; its highest-weight alias, $\Delta/10 = 74.92$\,d ($w = 0.28$), corresponds to 0.73\,S⊕. It is among the most promising unresolved habitable-zone candidates in our sample and a priority for follow-up.

\subsection{Mono-transits: a separate and harder problem}

Mono-transits contribute to the habitable-zone census, but their numbers must be read differently, for three reasons.

First, the catalogue has no period for them \citep{Guerrero2021}, yet ExoFOP lists an insolation for almost all of them (e.g. 0.38\,S⊕ for TOI-2433.01 and 11\,S⊕ for TOI-125.04). The period assumed for that insolation is not documented. Kepler catalogues, by contrast, explicitly assigned single-transit candidates a period estimated from the transit duration and stellar radius, flagged with a negative sign \citep{Batalha2013}. We tested whether the TESS values follow a similar rule by inverting the catalogue insolation of the 105 mono-transits in the full catalogue with their TIC luminosity and mass. The implied periods span 7.5--757\,d (5th--95th percentiles), show no preferred default value, and bear no consistent relation to the period implied by the transit duration for a central circular orbit (their ratio spans 0.07--48), to the number or time span of the observed sectors, to a fixed semi-major axis, or to the detection pipeline. The catalogue equilibrium temperatures, in contrast, follow from the insolation for a Bond albedo of 0.3 in 99 per cent of cases. The origin becomes clear in the SPOC Data Validation (DV) reports. For TOI-4309.01, the catalogue values (6.38\,S⊕, 405\,K) are exactly those of the DV report of the Sectors 1--65 multi-sector search (6.379\,S⊕, 405.3\,K), computed for the period of that threshold-crossing event, 53.67\,d. In DV the semi-major axis is not fitted but derived from that period and the stellar mass, and the equilibrium temperature assumes an albedo of 0.3. For a single transit the event period is spurious: the three multi-sector searches of this star pair the transit with different features and return 87.22, 53.67 and 37.80\,d, while we resolve the period at 41.04\,d. The catalogue retains the insolation of one of these runs after the period is set to zero, and does not update it when later searches change. The mechanism is general. Of the 105 single-transit TOIs in the catalogue, 96 have SPOC DV reports, and for 87 of them (91 per cent) the catalogue insolation matches that of one of their DV reports to better than 1 per cent (89 to better than 5 per cent), computed for the period of the corresponding threshold-crossing event (e.g. 175.52\,d for TOI-6666.01 and 37.82\,d for TOI-125.04, which we find at 49.06\,d). All the mismatches, and the one target whose reports could not be retrieved, are TOIs alerted by QLP, whose stars were later searched by SPOC; every match is a TOI of SPOC origin. For QLP-alerted mono-transits, and for the nine without DV reports, the insolation presumably comes from the QLP fit, for which no equivalent public report exists. This explains why the mono-transits from multi-sector SPOC searches deviate most from the duration-based period. The values cannot be reproduced from the catalogue parameters; we treat them as uninformative, so that `placed in the habitable zone by the catalogue' has no well-defined meaning for a mono-transit. Neither the catalogue insolation nor a catalogue-period criterion such as that of \citet{Kaltenegger2021} can therefore classify mono-transits.

Second, their period is constrained only by the transit shape and by the absence of further transits. The standard approach in the literature converts the transit duration and impact parameter, together with an independent stellar density, into a period through Kepler's third law \citep{Winn2010,Wang2015,Kipping2018}. \citet{Osborn2016} developed this for K2 single transits, and \citet{Sandford2019} combined Gaia DR2 parallaxes with isochrone masses to sharpen the stellar density for K2 single transits. Gaps between visits make the period posterior highly multimodal \citep{Dholakia2020}, and the window function of the survey must be included as a prior, since a long period is less likely to produce a transit within the observed baseline \citep[e.g.][]{Vanderburg2018,Rizzuto2020}. Our continuous period grid with duration likelihood, geometric prior and per-epoch absences follows the same logic, but with the full TESS archive the absences exclude far more of the period range than in single-campaign K2 or single-sector TESS data.

Third, even so, most mono-transits keep wide period posteriors, and the prior controls their habitable-zone probability: the expected number in the conservative habitable zone over all mono-transits is 8.6 under prior A and 12.6 under prior B, against 6.3--6.8 for duo-transits. We therefore quote mono-transit values as upper envelopes. The practical route to their periods is the one taken for TOI-2298\,b: wait until TESS observes further transits, or recover them from ground-based archival surveys and targeted follow-up, as done for TESS single transits with KELT, WASP and NGTS \citep{Yao2019,Gill2020}, and with CHEOPS for duo-transits \citep{Osborn2022}. Machine-learning estimators of period and equilibrium temperature from a single transit are a promising way to rank mono-transits for such follow-up \citep{Javed2026}.

TOI-2298\,b shows what the mono-transit sample can yield: catalogued today with no period, and in 2020 with a 165.2-d period from a single transit \citep{Kaltenegger2021}, it has a habitable-zone probability of 0.98 in our analysis and, once resolved and validated, is the only new habitable-zone planet in this work.

\section{Four validated planets and one candidate}
\label{sec:planets}

We selected for full characterisation the five new small-planet periods not compromised by contamination: TOI-2298.01, TOI-4503.01, TOI-6075.01, TOI-5724.01 and TOI-4629.01. Four are statistically validated; TOI-4503.01 is not (Section~\ref{sec:4503}).

\subsection{Joint transit fits}
\label{sec:fits}

\begin{figure*}
\centering
\includegraphics[width=\textwidth]{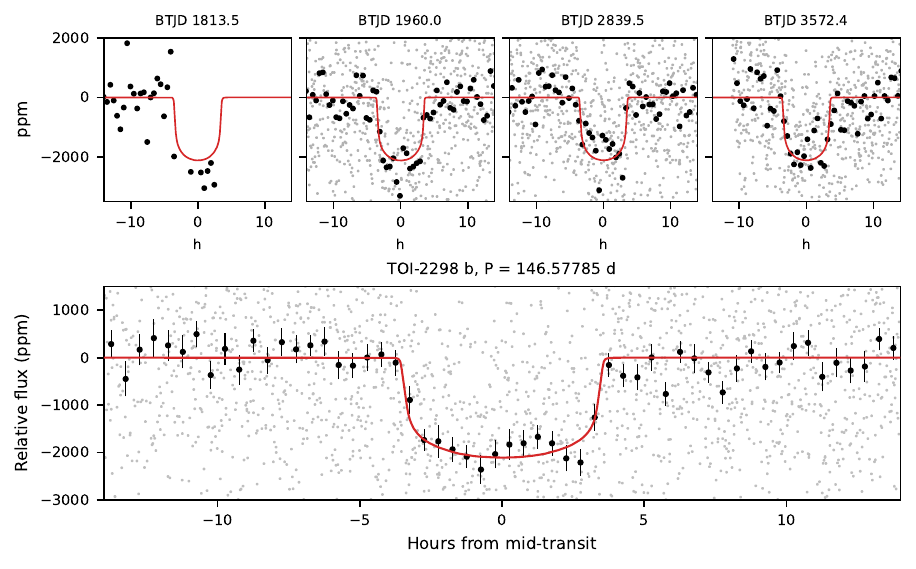}
\caption{TOI-2298\,b. Top: the four transits, with the \textsc{juliet} model (red); the Sector 18 data are TESS-SPOC 30-min full-frame-image fluxes, shown unbinned, and the other sectors SPOC 2-min fluxes binned to 30\,min. Bottom: all transits folded at $P = 146.57785$\,d, binned to 30\,min.}
\label{fig:toi2298}
\end{figure*}

\begin{figure*}
\centering
\includegraphics[width=\textwidth]{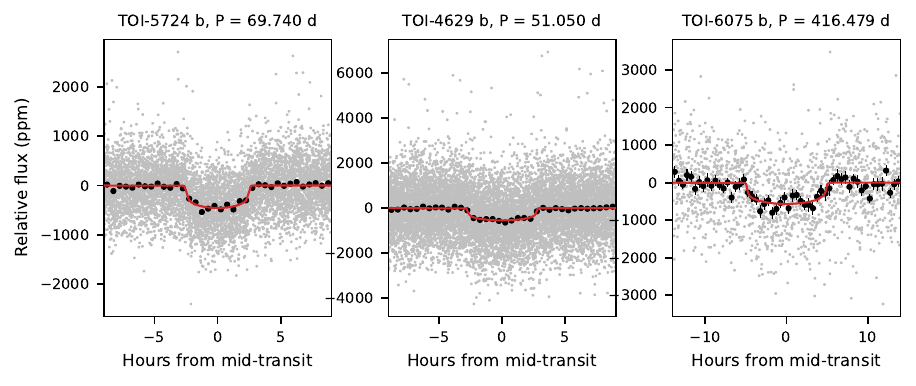}
\caption{Phase-folded transits of TOI-5724\,b, TOI-4629\,b and TOI-6075\,b with their \textsc{juliet} models, binned to 30\,min. TOI-6075\,b is folded at 416.48\,d; the 833-d solution is not excluded (Section~\ref{sec:fits}).}
\label{fig:validated}
\end{figure*}

We fitted all transits of each target with \textsc{juliet} \citep{Espinoza2019}, using \textsc{batman} \citep{Kreidberg2015} and \textsc{dynesty} \citep{Speagle2020} with 500 live points. Data within $\pm3$ durations of each transit were normalised by a quadratic baseline fitted out of transit. Each cadence (120, 200, 600 and 1800\,s) was treated as a separate instrument with its own flux offset and jitter, sharing quadratic limb-darkening coefficients in the $q_1$--$q_2$ parametrisation \citep{Kipping2013}; cadences longer than 120\,s were supersampled by a factor of 15. We sampled the radius ratio and impact parameter through $r_1$ and $r_2$ \citep{Espinoza2018}, fixed the eccentricity to zero and the dilution to unity, since PDCSAP fluxes are already corrected for crowding, and placed a Gaussian prior on $\rho_\star$ from the TIC mass and radius. Results are listed in Table~\ref{tab:fits} and shown in Figs~\ref{fig:toi2298} and~\ref{fig:validated}.

In all five fits the posterior stellar density agrees with the prior (e.g. 3190 against 3230\,kg\,m$^{-3}$ for TOI-2298), as expected if the period is correct and the transits occur on the catalogued star. As a stricter test we refitted TOI-2298.01, TOI-4503.01 and TOI-6075.01 with a log-uniform prior on $\rho_\star$ between 100 and $10^5$\,kg\,m$^{-3}$. The data alone constrain $\rho_\star$ weakly, through the usual degeneracy with the impact parameter (TOI-2298: $1330^{+1924}_{-671}$\,kg\,m$^{-3}$; TOI-4503: $196^{+2969}_{-53}$\,kg\,m$^{-3}$; TOI-6075: $2480 \pm {\sim}1700$\,kg\,m$^{-3}$), but in each case the Bayesian evidence favours the model with the stellar density, by $\Delta\ln Z = +1.3$, $+3.0$ and $+2.1$ respectively.

For TOI-6075.01 the two observed transits are separated by 832.96\,d and the alias 416.48\,d is not excluded by an absence. Fits with each period fixed give $\Delta\ln Z = \ln Z(416) - \ln Z(833) = +1.1$; the 833-d solution requires a higher impact parameter ($b = 0.73 \pm 0.08$ against $0.44 \pm 0.2$) to reproduce the 10.1-h duration. Including the geometric transit probability, which favours the shorter period by $2^{2/3}$, the odds are about 5:1 in favour of 416.48\,d. Only that solution predicts a transit on 2027 October 31 (BJD\,≈\,2461710.1).

\subsection{Difference-image centroids}
\label{sec:centroids}

For each transit we built a difference image (median out-of-transit flux, 0.75--2.5 durations from mid-transit, minus median in-transit flux), following the approach of \citet{Bryson2013}, from the SPOC target pixel files or $11 \times 11$-pixel TESScut cut-outs, and measured its flux-weighted centroid around the brightest difference pixel within 3 pixels of the target. Uncertainties combine a Monte Carlo propagation of the per-pixel noise with a systematic floor of 0.1 pixel (2.1″). Isolated offsets accompanied by a dipole (adjacent positive and negative difference pixels) were attributed to pointing motion during the event rather than to a contaminating source. For each neighbour that TRICERATOPS identifies as a possible host (Section~\ref{sec:fpp}) we measured the distance of the centroid from the target and from that neighbour.

\begin{figure*}
\centering
\includegraphics[width=0.92\textwidth]{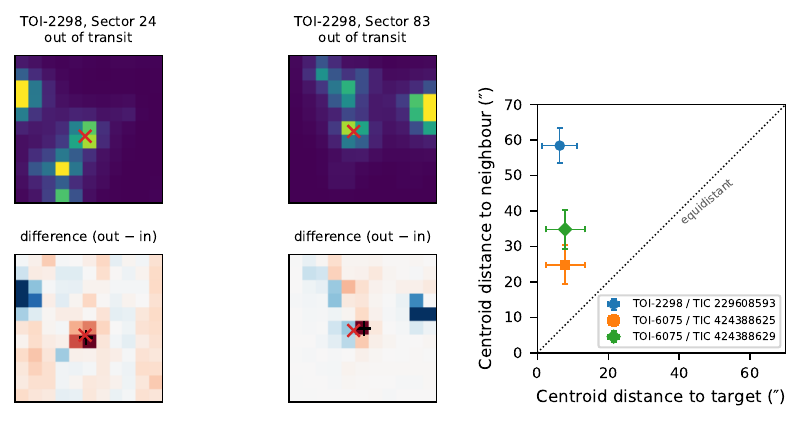}
\caption{Difference-image centroids. Left and centre: out-of-transit and difference images of TOI-2298 in Sectors 24 and 83 (red cross: target; black cross: flux-weighted centroid of the difference image, recomputed here for illustration). In Sector 24 the signal is centred on the target; in Sector 83 the adjacent positive and negative pixels form a motion dipole. A bright star at the edge of the cut-outs saturates the colour scale. Right: distance of the difference-image centroid from the target against its distance from each neighbour that TRICERATOPS identifies as a possible host (medians over events, with the typical uncertainty); all lie far from the equidistance line.}
\label{fig:centroids}
\end{figure*}

The transits of TOI-2298.01 are centred on the target in sectors 18, 24 and 56 (4.5--7.9″, $\leq 2.5\sigma$); the 17″ offset in sector 83 coincides with a motion dipole. The centroid is closer to the target than to the bright neighbour TIC\,229608593 (59.2″ away, $\Delta T = 0.28$\,mag) in all four events (median 6.3″ against 58.4″, $\sigma \approx 4.9$″). The difference image of sector 24, flagged on ExoFOP for background contamination, places the signal on the target pixel with no response at nearby stars. The two transits of TOI-6075.01 lie 9.0″ (with a motion dipole) and 6.5″ from the target; both neighbours flagged by TRICERATOPS are disfavoured, TIC\,424388625 (22.6″, $\Delta T = 2.2$\,mag) by $3.2\sigma$ and TIC\,424388629 (31.5″, $\Delta T = 5.8$\,mag) by $5.9\sigma$ per event. The events of TOI-5724.01 (three), TOI-4629.01 (four of five) and TOI-4503.01 (two) are compatible with the target (Fig.~\ref{fig:centroids}).

\subsection{Stellar properties, insolation and binarity}

With Gaia DR3 parameters (Table~\ref{tab:stars}), TOI-2298.01 receives $S = 0.62 \pm 0.03$\,S⊕ from a K dwarf with $\Teff = 4402$\,K and $L = 0.146$\,L☉. The conservative habitable zone at this temperature spans 0.97--0.28\,S⊕ \citep{Kopparapu2014}, so the planet lies near its centre, with a probability of 1.00 given the stellar uncertainties. TOI-4503.01 (0.20\,S⊕) and TOI-6075.01 (0.16 or 0.065\,S⊕) are colder than the outer edge; TOI-5724.01 and TOI-4629.01 receive about 19--20\,S⊕. All five hosts have RUWE\,$\leq 1.19$ and no multi-peak detections. As a control we applied the same procedure to TOI-4633, whose planet c has a published insolation of $1.56 \pm 0.2$\,S⊕; FLAME returns a luminosity twice that expected for the primary, consistent with blended light from its known stellar companion, while RUWE (1.15) shows no anomaly. RUWE is therefore insensitive to wide, long-period companions, which high-resolution imaging must exclude.

\subsection{Statistical validation}
\label{sec:fpp}

\begin{figure}
\centering
\includegraphics[width=\columnwidth]{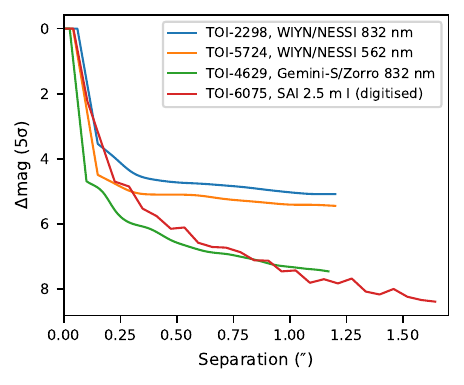}
\caption{High-resolution imaging contrast curves ($5\sigma$) used in the statistical validation. The SAI curve of TOI-6075 was digitised from the ExoFOP figure.}
\label{fig:contrast}
\end{figure}

We computed false-positive probabilities with TRICERATOPS \citep{GiacaloneDressing2020,Giacalone2021}, which considers all Gaia stars within 10 pixels and the background population from a TRILEGAL simulation \citep{Girardi2005}. Inputs were the SPOC pipeline apertures of up to six sectors containing transits, the phase-folded light curve within $\pm2$ durations binned to 10\,min, the period and depth from Section~\ref{sec:fits}, and the contrast curves of Section~\ref{sec:ancillary} (Fig.~\ref{fig:contrast}). Each calculation was repeated 20 times and we quote the mean and standard deviation. Where centroids exclude a neighbour as the source (Section~\ref{sec:centroids}), we removed the nearby-star scenarios of that neighbour (NTP, NEB, NEBx2P) and renormalised. We adopt the criteria of \citet{Giacalone2021}: FPP\,$< 0.015$ and NFPP\,$< 10^{-3}$ for validation.

All four planets meet both criteria (Table~\ref{tab:fpp}). TOI-5724\,b (FPP\,=\,$0.0011 \pm 0.0010$) and TOI-4629\,b ($0.0038 \pm 0.0022$) require no exclusions and have negligible NFPP. TOI-2298\,b (FPP\,=\,$0.0019 \pm 0.0021$, NFPP\,=\,$3 \times 10^{-7}$) and TOI-6075\,b (FPP\,=\,$0.0009 \pm 0.0007$, NFPP\,≈\,0) depend on the centroid-based exclusions: without them, TOI-2298\,b gives FPP\,=\,0.0029 and NFPP\,=\,$(1.1 \pm 0.5) \times 10^{-3}$, at the threshold, and TOI-6075\,b gives FPP\,=\,0.025 and NFPP\,=\,0.024. The contrast curves are decisive: without them the FPP values are 0.015--0.055, at or above the threshold, dominated by planets or eclipsing binaries on unresolved bound companions. TOI-6075\,b was validated with $P = 416.48$\,d; the 833-d solution is equally planetary.

\subsection{TOI-4503.01}
\label{sec:4503}

TOI-4503.01 has two transits of equal depth ($853 \pm 55$ and $764 \pm 49$\,ppm) separated by 710.74\,d, consistent centroids, a clean RUWE (0.94) and a stellar density favoured by the data. Every shorter alias is excluded by a single epoch at more than $15\sigma$, and the 11.5-h duration matches a ${\sim}2$-yr orbit around this K dwarf. Nevertheless, TRICERATOPS returns FPP\,≈\,1, dominated by eclipsing binaries at twice the period on an unresolved companion (61 per cent) or in the background (37 per cent), in which case the two events would be primary and secondary eclipses. The result persists with flat scenario priors (FPP\,=\,1.000, NFPP\,=\,0.11). Two shallow transits do not constrain the transit shape enough to separate these scenarios, and the geometric penalty of $a/R_\star \approx 414$ weighs against the planet. We therefore report TOI-4503.01 as a candidate with a resolved period. A contrast curve would address the companion scenarios, and its next transit, on 2026 November 22 at about 07:50\,UT ($\pm 45$\,min, duration ≈\,11.5\,h), offers a rare opportunity for follow-up.

\begin{table*}
\centering
\caption{Transit parameters (\textsc{juliet}, stellar-density prior, circular orbit). $T_{\rm eq}$ assumes zero albedo and full redistribution.}
\label{tab:fits}
\small
\setlength{\tabcolsep}{4pt}
\begin{tabular}{lccccc}
\toprule
Parameter & TOI-2298\,b & TOI-5724\,b & TOI-4629\,b & TOI-6075\,b & TOI-4503.01 \\
\midrule
$P$ (d) & $146.57785 \pm 0.00075$ & $69.74012 \pm 0.00019$ & $51.04974 \pm 0.00010$ & $416.479 \pm 0.012^\dagger$ & $710.740^{+0.005}_{-0.009}$ \\
$t_0$ (BJD$_{\rm TDB} - 2450000$) & $9399.7621 \pm 0.0037$ & $9675.6804 \pm 0.0015$ & $10102.7284 \pm 0.0030$ & $8794.755 \pm 0.011$ & $8523.8643 \pm 0.0086$ \\
$\Rp/R_\star$ & $0.0424 \pm 0.0015$ & $0.0228 \pm 0.0010$ & $0.0225 \pm 0.0008$ & $0.0221 \pm 0.0013$ & $0.0241 \pm 0.0018$ \\
$b$ & $0.28^{+0.21}_{-0.18}$ & $0.86 \pm 0.02$ & $0.66 \pm 0.06$ & $0.45^{+0.17}_{-0.28}$ & $0.54^{+0.10}_{-0.14}$ \\
$a/R_\star$ & $154^{+8}_{-12}$ & $55.5 \pm 3.7$ & $53.8 \pm 3.2$ & $292^{+20}_{-25}$ & $414^{+24}_{-29}$ \\
$i$ (°) & $89.90 \pm 0.08$ & $89.12 \pm 0.08$ & $89.30 \pm 0.11$ & $89.91 \pm 0.05$ & $89.93 \pm 0.02$ \\
$T_{14}$ (h) & $7.27^{+0.26}_{-0.22}$ & $5.37 \pm 0.12$ & $5.69 \pm 0.20$ & $10.1^{+0.7}_{-0.9}$ & $11.5^{+0.6}_{-0.5}$ \\
$\Rp$ (R⊕) & $3.04 \pm 0.29$ & $3.43 \pm 0.23$ & $2.76 \pm 0.16$ & $1.74 \pm 0.19$ & $2.09 \pm 0.19$ \\
$S$ (S⊕) & $0.62 \pm 0.03$ & $19.3 \pm 0.5$ & $19.9 \pm 0.5$ & $0.16 \pm 0.02$ & $0.201 \pm 0.007$ \\
$T_{\rm eq}$ (K) & 247 & 584 & 589 & 177 & 187 \\
\bottomrule
\multicolumn{6}{l}{$^\dagger$ Or 832.959\,d; the 416-d solution is favoured at about 5:1 (Section~\ref{sec:fits}).}
\end{tabular}
\end{table*}

\begin{table*}
\centering
\caption{Host stars of the characterised planets.}
\label{tab:stars}
\begin{tabular}{lccccccl}
\toprule
TOI & TIC & \Teff\ (K) & $L$ (L☉) & $M$ (M☉) & $R$ (R☉, TIC) & RUWE & Source of $L$, $M$ \\
\midrule
2298 & 229608594 & 4402 & 0.146 & 0.706 & $0.660 \pm 0.059$ & 1.11 & Gaia DR3 FLAME \\
5724 & 162631539 & 6012 & 2.316 & 1.143 & $1.375 \pm 0.070$ & 0.90 & Gaia DR3 FLAME \\
4629 & 304339227 & 5893 & 1.474 & 1.029 & $1.126 \pm 0.051$ & 1.00 & Gaia DR3 FLAME \\
6075 & 424388628 & 4192 & 0.146 & 0.659 & $0.725 \pm 0.066$ & 1.19 & TIC 8.2 \\
4503 & 167719618 & 5226 & 0.445 & 0.871 & $0.796 \pm 0.043$ & 0.94 & Gaia DR3 FLAME \\
\bottomrule
\end{tabular}
\end{table*}

\begin{table*}
\centering
\caption{Statistical validation (TRICERATOPS, 20 runs).}
\label{tab:fpp}
\small
\setlength{\tabcolsep}{4pt}
\begin{tabular}{lllcccl}
\toprule
Planet & Contrast curve & Neighbours excluded & FPP & NFPP & FPP, NFPP & Status \\
 & & by centroids & & & without exclusions & \\
\midrule
TOI-2298\,b & WIYN/NESSI 832\,nm & TIC 229608593 & $0.0019 \pm 0.0021$ & $3 \times 10^{-7}$ & 0.0029, $1.1 \times 10^{-3}$ & validated \\
TOI-5724\,b & WIYN/NESSI 562\,nm & none & $0.0011 \pm 0.0010$ & $< 10^{-60}$ & -- & validated \\
TOI-4629\,b & Gemini-S/Zorro 832\,nm & none & $0.0038 \pm 0.0022$ & $3 \times 10^{-11}$ & -- & validated \\
TOI-6075\,b & SAI 2.5\,m, $I$ (digitised) & TIC 424388625, TIC 424388629 & $0.0009 \pm 0.0007$ & $< 10^{-6}$ & 0.025, 0.024 & validated \\
TOI-4503.01 & none & none & ≈\,1 & 0.1--0.4 & -- & candidate \\
\bottomrule
\end{tabular}
\end{table*}

\section{Discussion}
\label{sec:discussion}

\subsection{TOI-2298\,b, a temperate sub-Neptune around a K dwarf}

\begin{figure}
\centering
\includegraphics[width=\columnwidth]{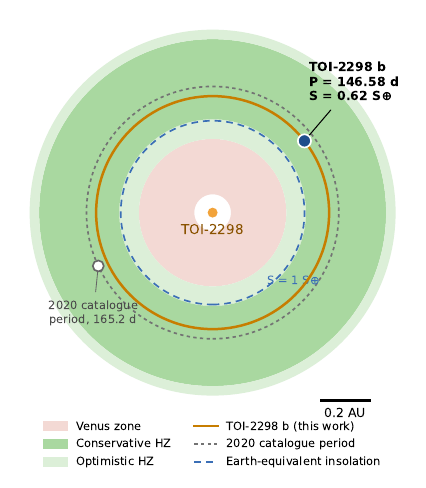}
\caption{Architecture of the TOI-2298 system, top-down and to scale, with Gaia DR3 stellar parameters ($L = 0.146$\,L☉, $M = 0.706$\,M☉, $\Teff = 4402$\,K). Shaded rings show the Venus zone \citep{Kane2014}, and the conservative and optimistic habitable zones \citep{Kopparapu2014}. The solid orbit is that of TOI-2298\,b ($P = 146.578$\,d, $a = 0.48$\,au); the dotted orbit corresponds to the 165.2-d period catalogued in 2020 from a single transit \citep{Kaltenegger2021}; the dashed circle marks Earth-equivalent insolation.}
\label{fig:architecture}
\end{figure}

TOI-2298.01 was first flagged as a possible habitable-zone planet by \citet{Kaltenegger2021}, on the basis of a single transit and a catalogue period of 165.2\,d, with a recommendation for further investigation. The four transits found here fix its period at 146.578\,d and confirm that it lies in the conservative habitable zone (Fig.~\ref{fig:architecture}), a stronger statement than the empirical (recent Venus--early Mars) limits used by \citet{Kaltenegger2021}, which correspond to our optimistic zone. Scaling their insolation of 0.48\,S⊕ at 165.2\,d to the true period gives 0.56\,S⊕ with the same TIC-8 parameters, and our Gaia DR3 value is 0.62\,S⊕. TOI-2298\,b combines a radius of 3.0\,R⊕, an equilibrium temperature of about 250\,K (zero albedo, full redistribution) and a position near the centre of the conservative habitable zone. Its size implies a substantial volatile envelope, so `habitable zone' describes its irradiation, not its surface. It belongs to the small class of transiting temperate sub-Neptunes with known periods, of which K2-18\,b is the best-studied member \citep{Montet2015}. Its JWST spectrum shows CH$_4$ and probably CO$_2$ \citep{Madhusudhan2023}, while the proposed sulphur-bearing biosignatures remain statistically unsupported in independent analyses \citep{Luque2025,Stevenson2025}, which illustrates both the scientific interest of these planets and the difficulty of interpreting their atmospheres.

Its atmospheric prospects are modest. Using the mass--radius relation of \citet{ChenKipping2017}, $M \approx 9.5$\,M⊕, the transmission spectroscopy metric of \citet{Kempton2018} is TSM\,≈\,21, against ≈\,55 for K2-18\,b computed in the same way and below the threshold of 90 recommended for priority targets. For a hydrogen-dominated atmosphere the scale height is about 90\,km, giving features of roughly 15\,ppm per scale height; detection would require several transits, of which a 146.6-d period allows at most two or three per year. A mass measurement is demanding but feasible: the expected radial-velocity semi-amplitude is $K \approx 1.5$\,m\,s$^{-1}$. At $\delta = +74$° the star is accessible only from the north (e.g. HARPS-N, KPF, MAROON-X).

The immediate priority is to recover the next transit, predicted for 2027 February 15 at 08:27\,UT\,$\pm$\,16\,min (BJD $2461451.852 \pm 0.011$), with a depth of about 1800\,ppm and a duration of 7.3\,h. The star is circumpolar from northern mid-latitudes, and a ground-based detection would confirm the period independently and settle the sector-24 contamination with seeing-limited resolution.

\subsection{Small planets on orbits longer than a year}

TOI-6075\,b (1.74\,R⊕, $P = 416.48$ or 832.96\,d) and the candidate TOI-4503.01 (2.09\,R⊕, $P = 710.74$\,d) belong to a nearly empty region of the transiting-planet parameter space. The NASA Exoplanet Archive (queried 2026 October) lists a single confirmed transiting planet smaller than 2\,R⊕ with $P > 1$\,yr, Kepler-452\,b (1.63\,R⊕, 384.8\,d), whose planetary nature has itself been questioned \citep{Mullally2018}. TOI-6075\,b is therefore the second validated planet in this regime and the first found by TESS, whichever of its two periods is correct. Both orbit K dwarfs and are colder than the habitable zone. Their periods were obtainable only because TESS has now observed their fields in several years. For TOI-6075\,b the transit of 2027 October 31 will decide between the two periods; for TOI-4503.01, the transit of 2026 November 22 is the last opportunity before 2028.

\subsection{Implications for the TOI catalogue and for demographics}

The catalogue period of a long-period TOI reflects the data available when it was alerted, not the current archive. In our sample, eight of the twelve published periods we recover differ from the catalogue value, and at least 15 further objects have new periods established here. Studies that select habitable-zone candidates from the catalogue, or that use catalogue periods of long-period TOIs in occurrence-rate or architecture analyses, therefore inherit errors that are systematic, not random: the catalogue period is always the longest alias, so it biases the population towards long periods and low insolation around FGK stars. The gain from additional data is already measurable. With data through Sector 62, \citet{Bass2025} found that almost every TOI with a catalogue period above 500\,d was a duo-transit with no data at its aliases; with 45 more sectors (through Sector 107), our analysis of a sample dominated by such objects resolves 28 periods and fixes 8 more by exclusion. A full-archive alias check of the kind presented here, repeated as TESS observes, would remove this bias at negligible cost.

\subsection{Limitations}

The method is limited by four factors. First, the selection on $\Pcat \geq 200$\,d misses duo-transits with shorter separations. Second, mono-transit periods remain prior-dominated unless the field has extensive coverage, which makes their habitable-zone probabilities upper envelopes. Third, 21″ pixels limit the centroid tests to neighbours beyond about 10″; closer companions must be excluded by imaging, and RUWE does not detect wide binaries, as the TOI-4633 control shows. Fourth, the validation of TOI-2298\,b and TOI-6075\,b depends on centroid-based neighbour exclusions, and that of TOI-6075\,b on a digitised contrast curve; we report the values without these inputs so that readers can judge their weight. Statistical validation of long-period signals at modest signal-to-noise is also exposed to instrumental false alarms, which TRICERATOPS does not model \citep{Mullally2018}; for TOI-6075\,b, whose two transits are each detected individually and agree in depth, duration and centroid, the transit of 2027 October 31 will provide the independent detection that such cases ultimately require. TRICERATOPS priors are calibrated on planet occurrence at periods much shorter than two years, which penalises candidates such as TOI-4503.01 and limits statistical validation at the longest periods.

\subsection{Prospects}

Continued TESS observations, the CHEOPS mission and, from 2027, PLATO will test the predictions of Table~\ref{tab:fits} and resolve the OPEN objects of this census. The SES grids, trial-period weights and vetting products for all 187 TOIs are released with this paper so that the analysis can be repeated as new sectors become available.

\section{Conclusions}

We have reanalysed the 187 TOIs catalogued as long-period duo- or mono-transits with the full public TESS archive. Our main conclusions are:
\begin{enumerate}
\item A flux-based soft-veto alias analysis, with tests for global coherence, depth coherence, data edges and contamination by other planets in the system, recovers 12 published periods blindly and correctly declines to choose when the data contain no alias information.
\item The archive resolves the period of 28 TOIs and constrains 8 more by exclusion. After visual vetting, 15 periods are new, 7 of them for small ($\Rp < 4$\,R⊕) candidates; two of these are probably contaminated by neighbours.
\item Of the 28 duo-transits that the catalogue periods place in the conservative habitable zone (criterion of \citealt{Kaltenegger2021}), only about two are expected to lie there, both giant planets, and none of the 12 small ones; those resolved are aliases of much hotter planets. Across the whole catalogue the problem is confined to these long-period objects, which make up three quarters of the 43 TOIs placed in the conservative habitable zone by their catalogue periods; for mono-transits the answer is prior-dependent and the catalogue insolation itself is of uncertain origin. Conversely, habitable-zone planets can hide behind missing or incorrect catalogue periods: three published M-dwarf planets are recovered blindly, and TOI-2298\,b is a new one.
\item TOI-2298\,b, TOI-5724\,b, TOI-4629\,b and TOI-6075\,b are statistically validated (FPP\,$\leq 0.004$, NFPP\,$< 10^{-3}$) with speckle contrast curves and, for two of them, centroid-based neighbour exclusions.
\item TOI-2298\,b is a $3.04 \pm 0.29$\,R⊕ sub-Neptune with $P = 146.57785 \pm 0.00075$\,d and $S = 0.62 \pm 0.03$\,S⊕, in the conservative habitable zone of a K dwarf. Its next transit occurs on 2027 February 15.
\item TOI-6075\,b (1.74\,R⊕, $P = 416.48$\,d or 832.96\,d) is, after Kepler-452\,b, the only validated transiting planet smaller than 2\,R⊕ with an orbit longer than a year, and the first found by TESS; the candidate TOI-4503.01 (2.09\,R⊕, $P = 710.74$\,d) lies just above that size.
\end{enumerate}

Table~\ref{tab:catalogue} gives, for each of the 187 TOIs, its probability of lying in the habitable zone after marginalising the period aliases: it complements the habitable-zone list of \citet{Kaltenegger2021}, which by construction covers cool stars whose habitable-zone periods allow three transits in the TESS prime mission, by covering the long-period TOIs that such a list cannot reach, and plays for these candidates a role similar to that of the Kepler list of \citet{Kane2016}. Catalogue periods of long-period TOIs should be treated as provisional, and periodic full-archive reanalyses should accompany their use in demographic and habitability studies.

\section*{Acknowledgements}

I thank Daniel Fabrycky and Dillon Bass for comments on the preprint, which prompted the common-mode tests of Section~\ref{sec:artefacts}. This paper includes data collected by the TESS mission, which are publicly available from the Mikulski Archive for Space Telescopes (MAST). Funding for the TESS mission is provided by the NASA Explorer Program. This research has made use of the Exoplanet Follow-up Observation Program (ExoFOP; DOI: 10.26134/ExoFOP5) website, operated by the California Institute of Technology under contract with NASA under the Exoplanet Exploration Program, and of the NASA Exoplanet Archive. This work has made use of data from the European Space Agency mission Gaia, processed by the Gaia Data Processing and Analysis Consortium. We thank the TFOP observers who obtained and shared the high-resolution imaging (WIYN/NESSI, Gemini-South/Zorro, SAI 2.5-m speckle) and TRES spectra used here. The large language model Claude Opus 5.5 (Anthropic) was used to assist in writing and debugging the analysis code, in literature searches, and in drafting and language editing of the manuscript text. The author designed and directed the research, ran all analyses, reviewed and verified all code, results and text, and takes full responsibility for the content.

\section*{Data Availability}

All TESS data are public at MAST. The full classification of the 187 TOIs (Table~\ref{tab:catalogue}) is available as an ancillary file (\texttt{table2\_full.csv}) with the arXiv version of this article. The per-object trial-period weights, the vetting figures, the \textsc{juliet} posteriors and the TRICERATOPS scenario tables will be released in a public repository, together with the analysis code, upon publication of this article. Until then they are available from the author on reasonable request.


\section*{Supporting Information}

The full version of Table~\ref{tab:catalogue} is available as an ancillary file (\texttt{table2\_full.csv}) with the arXiv version of this article.

\noindent\textbf{Table~\ref{tab:catalogue}.} Full version: classification of the 187 TOIs: type in the catalogue and as analysed (duo or mono), state, a status summary, catalogue period, highest-weight period and its weight, harmonic-family weight under priors A and B, insolation at the best period and as listed by ExoFOP, radius, host \Teff, number of new detections and of absences at the best period, probability of lying in the conservative and optimistic habitable zones under priors A and B, and notes (machine-readable CSV).

\bsp
\label{lastpage}
\end{document}